\documentclass[aip,pop,amsmath,amssymb,showpacs,reprint,floatfix,lengthcheck]{revtex4-1}
\usepackage[colorlinks,linkcolor=blue,citecolor=blue,bookmarks,pdfstartview=FitH]{hyperref}
\usepackage{graphicx}
\usepackage{dcolumn}
\usepackage{bm}
\usepackage{url}
\usepackage{amssymb}
\usepackage{enumerate}
\usepackage{enumitem}

\begin{document}

\preprint{AIP/123-QED}

\title{\textcolor{blue}{Lower hybrid destabilization of trapped electron modes in tokamak and its consequences for anomalous diffusion}}

\author{A. Kuley}
\email{animesh47@gmail.com}
\affiliation{Department of Physics,Indian Institute of Technology Delhi, New Delhi-110016, India.}
\author{C. S. Liu}
\affiliation{Deaprtment of Physics, University of Maryland, College Park, MD 20742, USA.}
\author{V. K. Tripathi}
\affiliation{Department of Physics,Indian Institute of Technology Delhi, New Delhi-110016, India.}
\date{\today}

\begin{abstract}
Parametric coupling of  lower hybrid pump wave with  low frequency collisionless/weakly collisional trapped electron drift wave, with frequency lower than the electron bounce frequency is studied. The coupling produces two lower hybrid sidebands. The sidebands beat with the pump to exert a low frequency ponderomotive force on electrons that causes a frequency shift in the drift wave, leading to the growth of the latter. The short wavelength modes are destabilized and they enhance the anomalous diffusion coefficient.

\end{abstract}

\maketitle

\section{Introduction}
Drift waves driven by trapped electrons, both dissipative trapped electron modes (DTEM) and collisionless trapped electron modes (CTEM), are considered to be an important agent for anomalous transport in  tokamak 
\cite{kadomtsev1971trapped,RevModPhys.71.735,PhysRevLett.103.085004,deng2009properties,lin2009studies,diamond2009physics,wesson2011tokamaks,doyle2007plasma,conner1994survey,dimits2000comparisons,jenko2000electron,lin1998turbulent,chowdhury2009comprehensive}. 
The nonlinearity associated with the trapped electron modes (TEM) has been extensively investigated theoretically \cite{gang1990nonlinear,gang1991kinetic,hahm1991weak,hahm1991nonlinear,beer1996bounce}. These microinsatbities are normally investigated using computer codes, e.g., gyro-kinetic code  GTC \cite{PhysRevLett.103.085004,deng2009properties}, GYRO \cite{lin2009studies,kinsey2006effect}, GS2 \cite{ernst2004role}, EM-GLOGYSTO \cite{chowdhury2009comprehensive}.
The TEM driven turbulence and transport have also been studied experimentally in some tokamaks such as Alcator C-Mod \cite{ernst2004role}, Axially Symmetric Divertor Experiment (ASDEX) upgrade \cite{PhysRevLett.86.5498}, and DIII-D \cite{deboo2005search}.
 
Recent experiments in Alcator C- Mod \cite{PhysRevLett.102.035002,rice2009observations} reported strong modification to toroidal rotation profiles in the core region $(0<r/a<0.4)$ induced by lower hybrid current drive (LHCD). The change in the radial electric field produced by the LHCD makes a nonambipolar radial current, charging the plasma negatively with respect to its pre lower hybrid (LH) state. This appears due to resonant trapped electron pinch i.e., the canonical angular momentum absorbed by the resonant trapped electrons while interacting with the lower hybrid waves and experiencing a faster inward drift than the ions in the core. Liu \textit{et al}., \cite{PhysRevLett.26.621} have developed an elegant theoritical formalism for radial, cross-field diffusion due to the nonconseravation of azimuthal angular momentum in an axisymmetric toroidal system, which appears due to the electric field component along the magnetic field lines of force. 

The lower hybrid waves launched into a tokamak by a phased array of wave guides and propagating towards the center in a well defined resonance cone are known to excite parametric instabilities. The parametric coupling to ion cyclotron mode and quasi-mode has been found to be  prominenet in high density tokamak. The lower hybrid wave spectrum thus generated has significant influence over lower hybrid current drive. The four wave parametric coupling of lower hybrid pump wave to drift waves has also been recognized to be important. Liu and Tripathi \cite{liu1980stabilization} explained the supression of drift waves by four wave parametric process. The $\textbf{E}\times \textbf{B}$ electron drift due to a lower hybrid pump wave of finite wave number beats with the density perturbation associated with the drift wave to produce sideband nonlinear currents that drive lower hybrid waves at lower and upper sideband frequencies. The sideband waves couple with the pump to exert a ponderomotive force that causes frequency shift in the eigen frequency of the drift wave. When this frequency shift overcomes the frequency shift due to finite Larmor radius effects the drift wave is stabilized. The lower hybrid pump with wave number greater than drift wave numbers was shown to stabilize the entire spectrum of drift wave when the pump amplitude exceeds a threshold value. Praburam \textit{et al}., \cite{praburam1988lower} developed a nonlocal theory of this process in a cylindrical plasma column. Wong and Bellan \cite{wong1978enhancement} studied the lower hybrid wave destabilization of collisional drift wave in the Princeton L-3 device. Redi \textit{et al}. \cite{redi2005microturbulent}, have analyzed linear drift mode stability in Alcator C- Mod with radio frequency heating, using GS2 gyrokinetic code, and shown that  ion temperature gradient (ITG) and electron temperature gradient (ETG) modes are unstable outside the barrier region and not strongly growing in the core; in the barrier region ITG/TEM is only weakly unstable for experimental profiles which have been modified by ion cyclotron radio frequency heating.

In a large aspect ratio tokamak, a trapped electron population exists in a fraction of velocity space given by $\delta\lambda\sim\sqrt{\epsilon}$, where $\lambda$ is the paricle's pitch angle and $\epsilon=r/R$ is the inverse aspect ratio of a tokamak magnetic surface with minor and major radii, $r$ and $R$ respectively. The trapped particles complete many bounces in its magnetic well before suffering sufficient small angle collisions to detrap them. They influence the low frequency drift waves very significantly, and having a destabilizing influence on them. Recently we \cite{kuley2009stabilization} have carried out the  gyrokinetic formalism to study lower hybrid wave stabilization of ion temperature gradient driven modes, in which the longer wavelength drift waves are destabilized by the lower hybrid wave while the shorter wavelengths are suppressed. In this paper we study the four wave parametric coupling of a lower hybrid pump wave to trapped electron modes. 

The paper is organized as follows : in section II the basic model and linear response of pump and sidebands are described. Section III presents low frequency perturbation. Section IV contains the nonlinear response at sidebands, and growth rate have been calculated in Sec. V. Finally in section VI we have discuss the results.

\section{Basic model and Linear response of pump and sidebands}
 We consider a toroidal geometry with circular concentric magnetic surfaces, parametrized by the usual usual coordinates $(\textbf{r},\theta,\xi)$ represent the the minor radius, poloidal angle and the toroidal angle coordinates, and the magnetic field can be written as $\textbf{B}=B[\textbf{e}_{\xi}+(\epsilon/q)\textbf{e}_{\theta}]$, where
$B=B_{0}(1-\epsilon$ cos$\theta)$ is the magnitude of the magnetic field, $q$ is the safety factor, $\textbf{e}_{\xi}$ and $\textbf{e}_{\theta}$ are the unit vectors along toroidal and poloidal direction respectively. The equilibrium distribution functions for electrons and ions are Maxwellians i,e.,
\begin{eqnarray}
 f_{0e}^0=n(m/2\pi T_{e})^{3/2} exp(-mv^{2}/2T_{e}),\nonumber\\
 f_{0i}^0=n(m_{i}/2\pi T_{i})^{3/2} exp(-m_{i}v^{2}/2T_{i}),
\end{eqnarray}
where $m$, $m_{i}$ are the mass of electron and ion, $v$ is the velocity, and $T_{e}$, $T_{i}$ denote the electron and ion temperature respectively.\\

A high power lower hybrid wave is launched into the plasma with potential $\phi_{0}$, $\omega_{0}$ lies in the range $\Omega_{i}\ll\omega_{0}\ll\Omega_{c}$ and $\Omega_{i}$, $\Omega_{c}$ are the ion and electron cyclotron frequencies. The dispersion relation for the lower hybrid wave is $\omega_{0}^2=\omega_{LH}^2(1+(m_{i}/m) k_{0\parallel}^2/k_{0}^2)$. This wave imparts oscillatory velocity to electrons
\begin{eqnarray}
 \textbf{v}_{0\perp}&=&-\frac{m}{eB^2}\biggl[i\omega_{0}\nabla_{\perp}\phi_{0}-\frac{e}{m}\textbf{B}\times\nabla_{\perp}\phi_{0}\biggr],\nonumber\\
\textbf{v}_{0\parallel}&=&-\frac{e}{mi\omega_{0}}\nabla_{\parallel}\phi_{0},
\end{eqnarray}
The second term in $\textbf{v}_{0\perp}$ represents the $\textbf{E}\times \textbf{B}$ drift, which is much larger than the polarization drift (first term in the same equation).
This oscillatory velocity provides a coupling between the low frequency TEM mode of potential
\begin{equation}
 \phi=A e^{-i(\omega t-\textbf k\cdot\psi)},
\end{equation}
and lower hybrid wave sidebands of potential
\begin{equation}
  \phi_{j}=A_{j} e^{-i(\omega_{j} t-\textbf k_{j}\cdot\psi)},
\end{equation}
with $j=$1, 2, where $\omega_{1}=\omega-\omega_{0}$, $\omega_{2}=\omega+\omega_{0}$, $\textbf{k}_{1}=\textbf{k}-\textbf{k}_{0}$, and $\textbf{k}_{2}=\textbf{k}+\textbf{k}_{0}$
The linear response of electrons to the sidebands turns out to be
\begin{eqnarray}
 \textbf{v}_{j\perp}&=&-\frac{m}{eB^2}\biggl[i\omega_{j}\nabla_{\perp}\phi_{j}-\frac{e}{m}\textbf{B}\times\nabla_{\perp}\phi_{j}\biggr],\nonumber\\
\textbf{v}_{j\parallel}&=&-\frac{e}{mi\omega_{j}}\nabla_{\parallel}\phi_{j}.
\end{eqnarray}

\section{nonlinear low frequency response}

The pump and sidebands exert a low frequency ponderomotive force ${\textbf{F}_{P}}$ on electrons. ${\textbf{F}_{P}}$ has two components, perpendicular and parallel to the magnetic field. The response of elctrons to  ${\textbf{F}_{P\perp}}$ is strongly supressed by the magnetic field and is usually weak. In the parallel direction, the electrons can effectively respond to $F_{P\parallel}$, hence, low frequency nonlinearity at $\omega, \textbf{k}$ arises mainly through $F_{P\parallel}=-m\textbf{v}\cdot\nabla v_{\parallel}$.
The parallel ponderomotive force, using the complex number identity $Re \textbf{A} \times Re \textbf{B}=(1/2) Re [\textbf{A}\times \textbf{B}+ \textbf{A}^*\times \textbf{B}]$, for the background electrons can be written as 
\begin{eqnarray}
 F_{p\parallel}=ei k_{\parallel}\phi _{p}= -\biggl(\frac{m}{2}\biggr)\biggl[\textbf{v}_{0\perp}\cdot\nabla_{\perp}v_{1\parallel}+\textbf{v}_{1\perp}\cdot\nabla_{\perp}v_{0\parallel}\biggr]\nonumber\\
-\biggl(\frac{m}{2}\biggr)\biggl[\textbf{v}_{0\perp}^{*}\cdot\nabla_{\perp}v_{2\parallel}+\textbf{v}_{2\perp}\cdot\nabla_{\perp}v_{0\parallel}^{*}\biggr]\quad
\end{eqnarray}
Using Eqs.(2) and (5) and considering only the dominant $\textbf{E}\times\textbf{B}$ drift terms the ponderomotive potential $\phi_{p}$  in the limit $\omega<<k_{\parallel}v_{the}$, takes the form
\begin{eqnarray}
 \phi_{p\parallel}=-\frac{\phi_{0}\phi_{1}}{2B^2}\frac{\textbf{B}\times \textbf{k}_{0\perp}\cdot \textbf{k}_{1\perp}}{\omega_{0}\omega_{1}ik_{\parallel}}\biggl[\omega_{0}k_{1\parallel}-\omega_{1}k_{0\parallel}\biggr]\nonumber\\
-\frac{\phi_{0}^{*}\phi_{2}}{2B^2}\frac{\textbf{B}\times \textbf{k}_{0\perp}\cdot \textbf{k}_{2\perp}}{\omega_{0}\omega_{2}ik_{\parallel}}\biggl[\omega_{0}k_{2\parallel}-\omega_{2}k_{0\parallel}\biggr].
\end{eqnarray}
One may note $\phi_{p}$ is  maximum when $\textbf{k}_{\perp}$ and $\textbf{k}_{0\perp}$ are perpendicular to each other. The ponderomtive force on ions is weak, hence we ignore it and take the ion response to be linear.
The electron density perturbation due to $\phi$ and $\phi_{p}$ can be written in terms of electron susceptibility of $\chi_{e}$ as 
\begin{equation}
  \frac{\delta n_{e}}{n}=\frac{k^2\epsilon_{0}}{e}\chi_{e}(\phi+\phi_{p}),
\end{equation}
where as ion perturbation in terms of ion susceptibility $\chi_{i}$ as 
\begin{equation}
  \frac{\delta n_{i}}{n}=-\frac{k^2\epsilon_{0}}{e}\chi_{i}\phi.
\end{equation}
Here n is the equlibrium electron density and $\epsilon_{0}$ is the free space permittivity. For the ions, neglecting collisions, longitudinal motion and cross filed guiding center drifts one can write 
\begin{equation}
\chi_{i}=\frac{2\omega_{pi}^2}{k^2v_{thi}^2}\biggl[1-(1-\frac{\omega_{i}^{*}}{\omega})I_{0}e^{-b}-\eta_{i}\frac{\omega_{i}^{*}}{\omega}b(I_{0}-I_{1})e^{-b}\biggr]. 
\end{equation}
where $b=(k_{\perp}\rho_{i})^2$, $\rho$=$v_{th}/\omega_{c}$, $I_{0}$ and $I_{1}$ are the modified Bessel functions of zero and first order, respectively, $\omega_{i}^{*}$ is the ion diamagnetic drift frequency, $\textbf{k}_{\perp}$ is the perpendicular wave number, $\textbf{k}_{\perp}=k_{\theta}\textbf{e}_{\theta}+k_{\textbf{r}}\textbf{e}_{\textbf{r}}$.\\
 
For the electron susceptibility we consider two cases.
\subsection{Collisionless TEM Mode}
In this case susceptibility can be taken from Ref.\cite{gang1991kinetic}
\begin{equation}
\chi_{e}=\frac{2\omega_{p}^2}{k^2v_{the}^2}\biggl[1+\biggl(\frac{\epsilon}{2}\biggr)^{1/2}\biggl(1-\frac{\omega_{e}^*}{\omega}\biggr)\triangle ln\frac{x_{t}}{\triangle}\biggl(g_{n}\sqrt{\frac{i\mu}{\pi}}\biggr)\biggr].
\end{equation}
where $\triangle$ is the separation of adjacent mode rational surfaces for fixed toroidal mode number, $\triangle=1/k_{\theta}\hat{s}$ (it also signifies the trapped electron layer width, which demarks the region in which the trapped electron response is significant), $x_{t}=\sqrt{L_{n}/L_{s}} \rho$ represents the turning point width, $L_{s}$, $L_{n}$, and $\rho$ are the magnetic shear length, equilibrium density scale length and ion Larmor radius and for the collisionless regime $(\omega_{De}<\omega<\omega_{b})$
\begin{equation}
 Im(g_{n})=2\sqrt\pi\biggl(\frac{\omega}{\omega_{De}}\biggr)^{3/2}e^{-\omega/\omega_{De}}
\end{equation}
with   $\omega_{De}=L_{n}/R \omega_{e}^{*}$, and $\omega_{e}^{*}$ is the elctron diamagnetic frequency.\\
Using the Eqs.(10) and (11) in the Poisson's equation, we obtain
\begin{equation}
 \varepsilon\phi=-\chi_{e}\phi_{p},
\end{equation}
where $\varepsilon=1+\chi_{i}+\chi_{e}$, 

\subsection{Weakly Collisional TEM Mode}
In the low collisionality '\textit{banana}' regime $\nu_{*e}=\nu_{e}/\omega_{be}\epsilon\ll$1, where $\nu_{e}$ is the $90^{0}$ coulomb collision frequency and $\omega_{be}=\epsilon^{1/2}v_{the}/Rq$ is the typical bounce frequency of trapped electrons, electron suscetibility and can be written as \cite{connor2006stability} 
\begin{eqnarray}
\chi_{e}^c=\frac{2\omega_{p}^2}{k^2v_{the}^2}\biggl\{1-\frac{2\sqrt{2\epsilon}}{\pi}\biggl(1-\frac{\omega_{e}^{*}}{\omega^c}\biggr)\nonumber\\
+\frac{2\sqrt{2\epsilon}\Gamma(3/4)}{\pi^{3/2}}(1+i)\sqrt{\frac{\nu_{the}}{\omega^c\epsilon}}\biggl[1-\frac{\omega_{*}^{e}}{\omega^c}\biggl(1-\frac{3}{4}\eta_{e}\biggr)\biggr]\biggr\}.
\end{eqnarray}
where $\nu_{the}$ is the collision frequency at thermal speed and we have neglected a small population of low energy elctrons which are highly collisional. 
Using the Eqs.(10) and (14) in the Poisson's equation  we obtain
\begin{equation}
 \varepsilon^{c}\phi=-\chi_{e}^{c}\phi_{p},
\end{equation}
where $\varepsilon^{c}=1+\chi_{i}+\chi_{e}^{c}$, and $\chi_{e}^{c}$ is the electron (weakly collisional) 

\section{NONLINEAR RESPONSE  AT THE SIDEBANDS}
The density perturbation at $(\omega, \textbf{k})$ couples with the oscillatory velocity of electrons, $\textbf{v}_{0}$, to produce nonlinear density perturbations at sideband frequencies. Solving the equation of continuity, 
\begin{equation}
 \frac{\partial}{\partial t}n_{1}^{NL}+\nabla\biggl(\frac{\delta n_{e}}{2}\textbf{v}_{0}^*\biggr)=0,
\end{equation}
one obtains 
\begin{equation}
 n_{1}^{NL}=\frac{\delta n_{e}}{2\omega_{1}}(\textbf{k}_{1}\cdot\textbf{v}_{0}^*).
\end{equation}
Similarly for the upper sideband the nonlinear density perturbation can be written as
\begin{equation}
 n_{2}^{NL}=\frac{\delta n_{e}}{2\omega_{2}}(\textbf{k}_{2}\cdot\textbf{v}_{0}).
\end{equation}
  Using Eqs. (17) and (18) in the Poisson's equation for the sideband waves, we obtain
\begin{eqnarray}
\varepsilon_{1}\phi_{1}=\frac{k^2}{k_{1}^2}\biggl(1+\chi_{i}\biggr)\frac{\textbf{k}_{1}\cdot{\textbf{v}}_{0}^*}{2\omega_{1}}\phi,\nonumber\\
\varepsilon_{2}\phi_{2}=\frac{k^2}{k_{2}^2}\biggl(1+\chi_{i}\biggr)\frac{\textbf{k}_{2}\cdot{\textbf{v}}_{0}}{2\omega_{2}}\phi,
\end{eqnarray}
where
\begin{eqnarray}
 \varepsilon_{1}=1+\frac{\omega_{p}^2}{\Omega_{c}^2}-\frac{\omega_{pi}^2}{\omega_{1}^2}\biggl(1+\frac{k_{1\parallel}^2}{k_{1}^2}\frac{m_{i}}{m}\biggr),\nonumber\\
 \varepsilon_{2}=1+\frac{\omega_{p}^2}{\Omega_{c}^2}-\frac{\omega_{pi}^2}{\omega_{2}^2}\biggl(1+\frac{k_{2\parallel}^2}{k_{2}^2}\frac{m_{i}}{m}\biggr),
\end{eqnarray}
are the dielectric functions at $(\omega_{1},\textbf{k}_{1})$, and $(\omega_{2},\textbf{k}_{2})$.
\section{growth rate}
The coupled equations (13) and (19) lead to the nonlinear dispersion relation,
\begin{equation}
 \varepsilon=-\frac{\chi_{e}(1+\chi_{i})}{(1+\frac{k_{0\perp}^2}{k_{\perp}^2})}\frac{k^2 U^2 {\sin}^2 \delta}{4\omega_{0}^2}\biggl[\frac{1}{ \varepsilon_{1}}+\frac{1}{ \varepsilon_{2}}\biggr],
\end{equation}
 where
$U=\lvert{k_{0}{{\phi}_{0}}}/{B}\rvert$ is the magnitude of $\textbf{E}_{0}\times \textbf{B}$ electron velocity, and $\delta$ is the angle between $\textbf{k}_{\perp}$ and $\textbf{k}_{0\perp}$. For ${\omega k_{0z}}/{\omega_{0}k_{z}} \ll1$, $k_{\perp}^2<k_{0\perp}^2$, $k_{z}^2<k_{0z}^2$ one may write 
\begin{equation}
 \frac{1}{\varepsilon_{1}}+\frac{1}{\varepsilon_{2}}\backsimeq\frac{1}{2(1+\omega_{p}^2/\Omega_{c}^2)}\frac{1-k_{\perp}^2k_{0z}^2/k_{z}^2k_{0\perp}^2}{(1-\omega_{LH}^2/\omega_{0}^2)}
\end{equation}

We simplify Eq. (21) in two different cases 
\subsection{Collisionless TEM Mode}
Writing $\omega=\omega_{r}+i\gamma$ ,with $\gamma\ll \omega_{r}$,  the real and imaginary parts  of Eq. (21)  gives 
\begin{equation}
 \omega_{r}=-\omega_{i}^*\frac{[I_{0}e^{-b}-\eta_{i}b(I_{0}-I_{1})e^{-b}]}{1-I_{0}e^{-b}+\frac{1}{T_{e}/T_{i}+P2\omega_{pi}^2/k^2v_{thi}^2}},\nonumber\\
\end{equation}
for $b<$1 
\begin{equation}
 \omega_{r}\backsimeq-\omega_{i}^*\biggl[\frac{1-b-\eta_{i}b}{b+\frac{1}{T_{e}/T_{i}+P2\omega_{pi}^2/k^2v_{thi}^2}}\biggr],
\end{equation}

\begin{equation}
 \gamma=-\frac{T_{i}}{T_{e}}\sqrt{\frac{\epsilon}{2}}\frac{(1-\frac{\omega_{e}^*}{\omega})\Delta ln\frac{x_{t}}{\Delta}Im(g_{n}\sqrt{\frac{i\mu}{\pi}})(1+P\chi_{i})}{1-I_{0}e^{-b}+\frac{1}{T_{e}/T_{i}+P2\omega_{pi}^2/k^2v_{thi}^2}}
\end{equation}
There are two regimes, for small $k_{\perp}$ regime, $P$ reduces, hence the drift wave frequency enhance and growth rate increases. For large $k_{\perp}$, $P$ become positive and hence the growthrate reduces, 
\begin{figure}
 \centering
 \includegraphics[width=9.3cm,height=7.5cm]{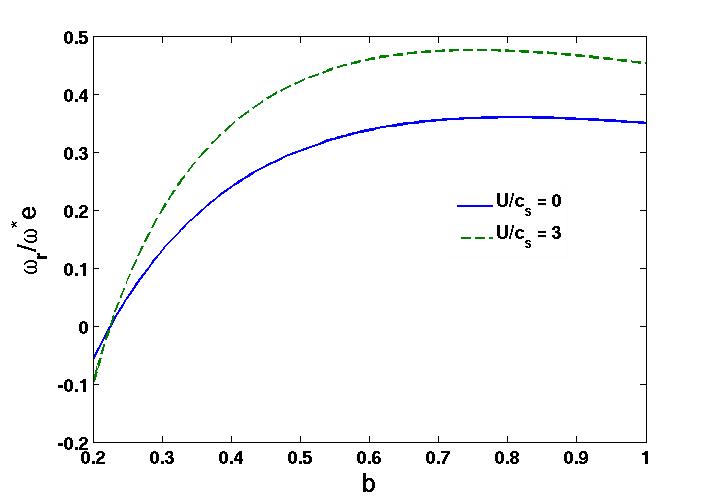}
 \caption{(Color online) Variation of normalized real frequency for collisionless TEM as a function of b for $\eta_{i}$=5, $R/L_{n}$ = 1.8}.
 \label{fig:1}
\end{figure}
where
\begin{eqnarray}
 P\simeq\frac{1-k_{\perp}^2k_{0z}^2/k_{z}^2k_{0\perp}^2}{2(1+\omega_{p}^2/\Omega_{c}^2)(1-\omega_{LH}^2/\omega_{0}^2)}\frac{k^2U^2}{4\omega_{0}^2(1+k_{0}^2/k^2)}.
\end{eqnarray}
\begin{figure}
 \centering
 \includegraphics[width=9.3cm,height=7.5cm]{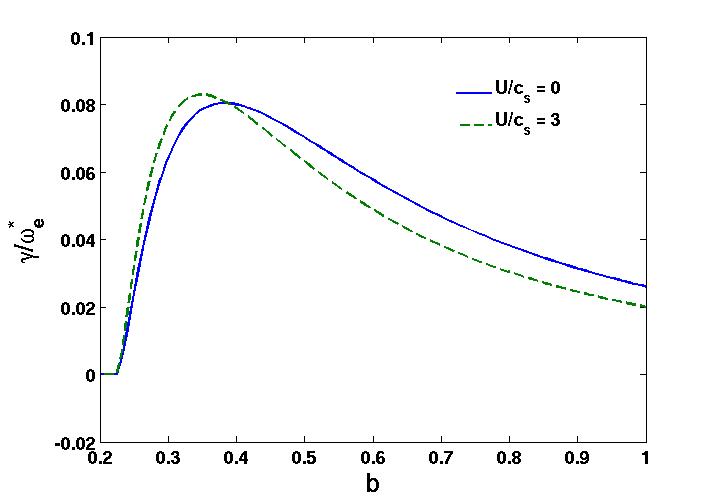}
 \caption{(Color online) Variation of normalized growthrate for collisionless TEM as a function of b for $\eta_{i}$=5, $R/L_{n}$ = 1.8}.
 \label{fig:2}
\end{figure}
 
\subsection{Weakly Collisional TEM Mode}
Writing $\omega^{c}=\omega_{r}^{c}+i\gamma^{c}$ ,with $\gamma^{c}\ll \omega_{r}^{c}$,  the real and imaginary parts  of Eq. (21)  gives 
\begin{figure}
 \centering
 \includegraphics[width=9.3cm,height=7.5cm]{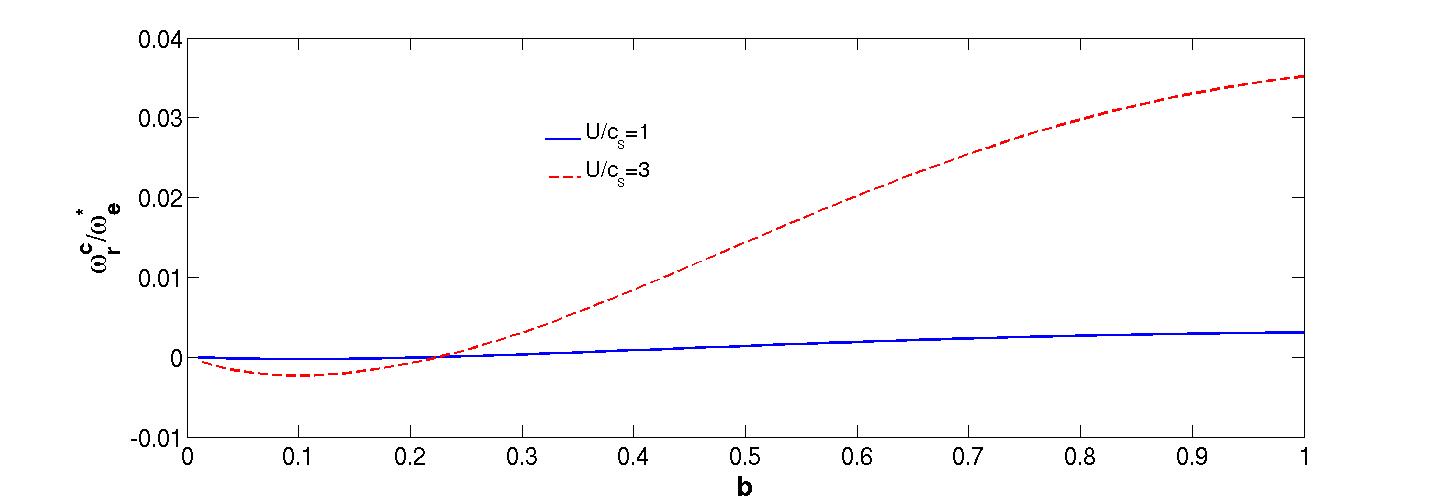}
 \caption{(Color online) Variation of normalized real frequency for weakly collisionl TEM as a function of b for two different values of lower hybrid amplitude $U/c_{s}$=1, and 3. Others paprameter are $\eta_{i}$=5, $R/L_{n}$ = 1.8, collisionality parameter $L_{n}\nu_{the}/v_{thi}=0.01$}.
 \label{fig:3}
\end{figure}
\begin{eqnarray}
 \omega_{r}^{c}[\frac{G}{\omega_{e}^*}-(1-\frac{3}{4}\eta_{e})(S+\frac{1}{P})]+\frac{G}{\Gamma(3/4)}\sqrt{\frac{\pi\epsilon}{\nu_{the}}}\sqrt{\omega_{r}^c}\nonumber\\
-G(1-\frac{3}{4}\eta_{e})=0,\nonumber\\
\gamma^c=\frac{\frac{G}{\Gamma(3/4)}\sqrt{\frac{\pi\epsilon}{\nu_{the}}}\sqrt{\omega_{r}^c}}{[\frac{G}{\omega_{e}^*}-(1-\frac{3}{4}\eta_{e})(S+1/P)]-\frac{G}{2\Gamma(3/4)}\sqrt{\frac{\pi\epsilon}{\nu_{the}\omega_{r}^c}}},
\end{eqnarray}
where
\begin{eqnarray}
G=2\frac{\omega_{pi}^2}{k^2v_{thi}^2}\omega_{i}^*\biggl\{I_{0}e^{-b}-\eta_{i}b(I_{0}-I_{1})e^{-b}\biggr\},\nonumber\\
S=1+2\frac{\omega_{pi}^2}{k^2v_{thi}^2}(1-I_{0}e^{-b}).
\end{eqnarray}
\begin{figure}
 \centering
 \includegraphics[width=9.3cm,height=7.5cm]{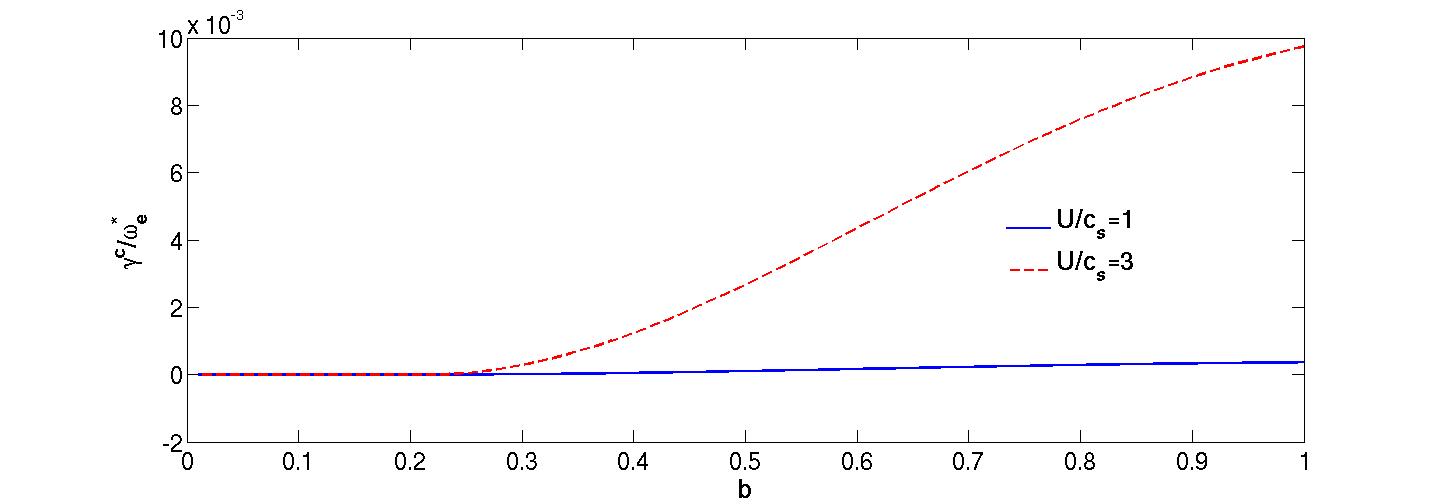}
 \caption{(Color online) Variation of normalised growth rate for weakly collisional TEM as a function of b for two different lower hybrid amplitude $U/c_{s}$=1, and 3. Others parameters are $\eta_{i}$=5, $R/L_{n}$ = 1.8, collisionality parameter $L_{n}\nu_{the}/v_{thi}=0.01$}..
 \label{fig:4}
\end{figure}

In order to have a numerical appreciation of results we consider the following set of parameters, corresponding to Alcator C-Mod tokamak \cite{PhysRevLett.102.035002}, a compact tokamak : major radius $R$ = 0.67 m, typical minor radius = 0.21 m, $r/a\sim<0.4$, background electron density $\sim$ $10^{20} m^{-3}$, electron temperature$\sim$ 2.5keV, ion temperature $\sim$ 1 keV, magnetic field $\sim$ 5T, frequency of the lower hybrid pump is 4.6 GHz, and the refractive index of the lower hybrid wave parallel to the magnetic field is $\sim$2, $R/L_{n}$=1.8,  $U/c_{s}$ =5, where $c_{s}$ is the ion sound speed. The value of $U/c_{s}=$  3 corresponds to lower hybrid power of 1.7 MW \cite{liu1984density}. One may mention that the range of lower hybrid power is typically $\sim$ 1 MW and looking for the increase of LH power to 2.0-2.4 MW in future.
\begin{figure}
 \centering
 \includegraphics[width=8cm,height=6.5cm]{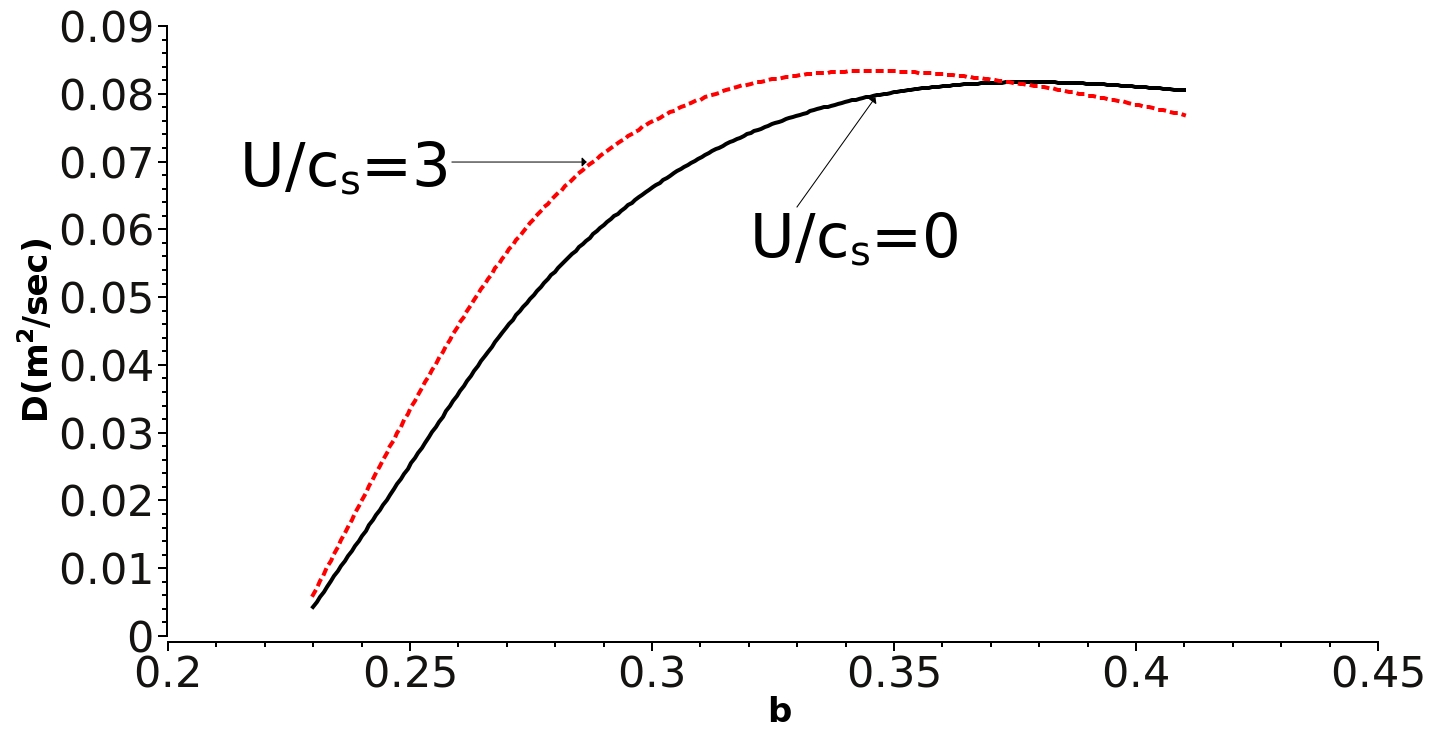}
 \caption{Variation of diffusion coefficient for collisionless TEM as a function of b for $\eta_{i}$=5, $R/L_{n}$ = 1.8, $\tau_{c}=20a/c_{s}$, $U/c_{s}=3$}.
 \label{fig:5}
\end{figure}

Figure 1 shows the progression of normalized wave frequency for electrostatic collisionless TEM mode as a function of b for different pump power $U/C_{s}$=0, and 3 which shows, lower hybrid pump amplitude have a significant effect on real frequency, while in case of growth rate of the collisionless TEM (cf. Fig.2) the lower hybrid amplitude has a very tiny effect on the destabilization of the drift wave, and significant effect on suppressing smaller wavelength drift wave.  

Figure 3 shows the progression of normalised wave frequency for electrostatic weakly collisional TEM mode as a function of b for different lower hybrid amplitude $U/c_{s}$=1, and 3, and collisionality parameter $L_{n}\nu_{the}/v_{thi}$=0.01. The longer wave length drift waves are stabilized by the lower hybrid pump wave, while the shorter wavelength get destabilized (cf. Fig.4)

Finally we consider the anomalous diffusion in an axisymmetric system, due to low-frequency, electrostatic instabilities, with charecterstic frequency lower than the mean bounce frequency of the trapped particles between the mirrors, the resulting resultant diffusion of the trapped particle is mainly due to the lack of conservation of the canonical angular momentum.

\begin{figure}
 \centering
 \includegraphics[width=9cm,height=8cm]{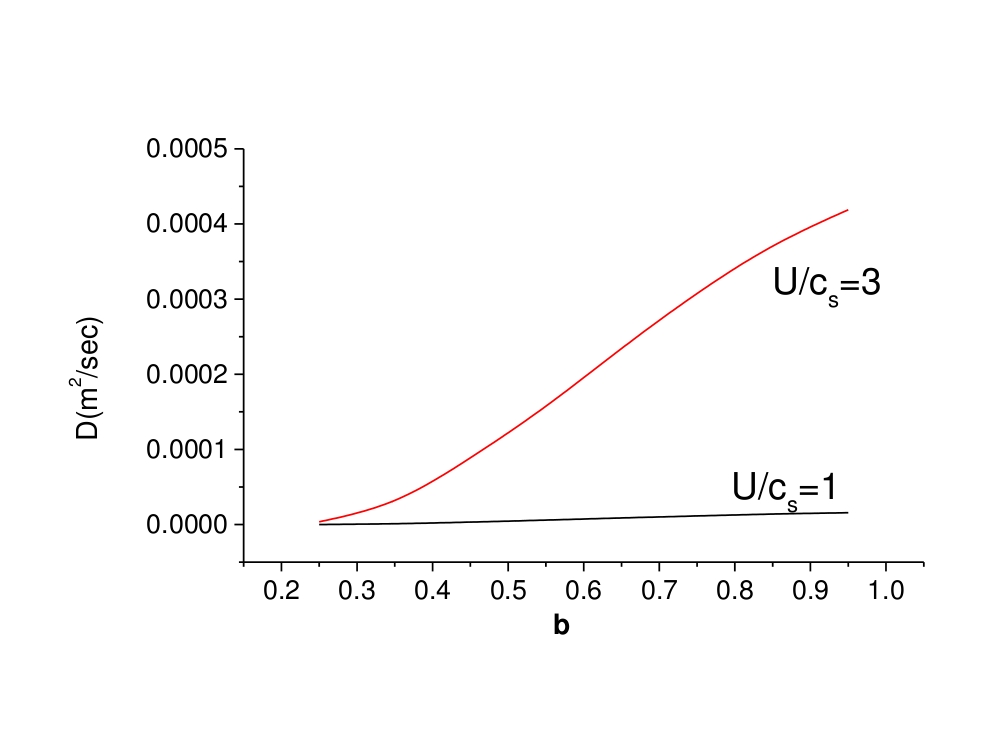}
 \caption{Variation of diffusion coefficient for weakly collisional TEM as a function of b for $\eta_{i}$=5, $R/L_{n}$ = 1.8, $\tau_{c}=20a/c_{s}$, collisionality parameter $L_{n}\nu_{the}/v_{thi}$=0.01}.
 \label{fig:6}
\end{figure}
The anomalous diffusion coefficient for the trapped particle can be written from Ref. \cite{PhysRevLett.26.621}
\begin{equation}
 D \approx  v_{\Omega}^2\tau_{c} \approx\frac{E_{\xi}^2}{B_{\theta}^2}\tau_{c}
\end{equation}
where $v_{\Omega}$ is the drift velocity of the trapped particle towards the magnetic axis, $\tau_{c}$ is the correlation time. The quantitate estimate of $\lvert {e\phi}/{T}\rvert^2 \sim ({\gamma}/{\omega^*})({1}/{k_{\perp}^2L_{n}^2})$.

In Figs.5 and 6 we have plotted the diffusion co-efficient for collisionless and weakly collisional TEM mode for different lower hybrid pump amplitude.
\section{Discussions}
 The anomalous diffusion of collisionless and weakly collisional trapped particles  due to low frequency modes is considered. In axisymmetric torus the diffusion of the trapped particles appear due to changes in the angular momentum (Ware Pinch \cite{PhysRevLett.25.15}). With the increasing of normalized lower hybrid pump amplitude it further destabilized the drift wave (cf. Figs. 3 and 4) by the parametric coupling of the pump and the sideband waves, which gives a significant role in diffusion of the trapped particle in the core region of the tokamak (cf. Figs. 5 and 6), as most recently observed in Alcator C-Mod \cite{PhysRevLett.102.035002,rice2009observations}. The inward diffusion of the trapped electrons in the presence of lower hybrid pump is quite significantly large compared with the weakly collisional trapped electron modes.
In the region of trapped particles the amplitude of pump wave has to be constant, which may be reasonable as long as pump frequency of the lower hybrid layer. The Lower hybrid wave - trapped particle mode interaction is localized in a parallel length of the order of the width of the phased array of the wave guides. However the drift wave mode structure extends far beyond this region, hence the pump effectiveness is may be significantly reduced. The trapped particle diffusion is primarily expected to be taken place in the LH resonance cone.
\section{Acknowledgement}
The Authors are greatful to  Prof. Zhiong Lin and Dr. Yong Xiao of University of California Irvine for valuable suggestions.

\nocite{*}
\bibliography{Kuley-10}

\providecommand{\noopsort}[1]{}\providecommand{\singleletter}[1]{#1}%
\begin{thebibliography}{33}%
\makeatletter
\providecommand \@ifxundefined [1]{%
 \@ifx{#1\undefined}
}%
\providecommand \@ifnum [1]{%
 \ifnum #1\expandafter \@firstoftwo
 \else \expandafter \@secondoftwo
 \fi
}%
\providecommand \@ifx [1]{%
 \ifx #1\expandafter \@firstoftwo
 \else \expandafter \@secondoftwo
 \fi
}%
\providecommand \natexlab [1]{#1}%
\providecommand \enquote  [1]{``#1''}%
\providecommand \bibnamefont  [1]{#1}%
\providecommand \bibfnamefont [1]{#1}%
\providecommand \citenamefont [1]{#1}%
\providecommand \href@noop [0]{\@secondoftwo}%
\providecommand \href [0]{\begingroup \@sanitize@url \@href}%
\providecommand \@href[1]{\@@startlink{#1}\@@href}%
\providecommand \@@href[1]{\endgroup#1\@@endlink}%
\providecommand \@sanitize@url [0]{\catcode `\\12\catcode `\$12\catcode
  `\&12\catcode `\#12\catcode `\^12\catcode `\_12\catcode `\%12\relax}%
\providecommand \@@startlink[1]{}%
\providecommand \@@endlink[0]{}%
\providecommand \url  [0]{\begingroup\@sanitize@url \@url }%
\providecommand \@url [1]{\endgroup\@href {#1}{\urlprefix }}%
\providecommand \urlprefix  [0]{URL }%
\providecommand \Eprint [0]{\href }%
\providecommand \doibase [0]{http://dx.doi.org/}%
\providecommand \selectlanguage [0]{\@gobble}%
\providecommand \bibinfo  [0]{\@secondoftwo}%
\providecommand \bibfield  [0]{\@secondoftwo}%
\providecommand \translation [1]{[#1]}%
\providecommand \BibitemOpen [0]{}%
\providecommand \bibitemStop [0]{}%
\providecommand \bibitemNoStop [0]{.\EOS\space}%
\providecommand \EOS [0]{\spacefactor3000\relax}%
\providecommand \BibitemShut  [1]{\csname bibitem#1\endcsname}%
\let\auto@bib@innerbib\@empty
\bibitem [{\citenamefont {Kadomtsev}\ and\ \citenamefont
  {Pogutse}(1971)}]{kadomtsev1971trapped}%
  \BibitemOpen
  \bibfield  {author} {\bibinfo {author} {\bibfnamefont {B.}~\bibnamefont
  {Kadomtsev}}\ and\ \bibinfo {author} {\bibfnamefont {O.}~\bibnamefont
  {Pogutse}},\ }\href@noop {} {\bibfield  {journal} {\bibinfo  {journal}
  {Nuclear Fusion}\ }\textbf {\bibinfo {volume} {11}},\ \bibinfo {pages} {67}
  (\bibinfo {year} {1971})}\BibitemShut {NoStop}%
\bibitem [{\citenamefont {Horton}(1999)}]{RevModPhys.71.735}%
  \BibitemOpen
  \bibfield  {author} {\bibinfo {author} {\bibfnamefont {W.}~\bibnamefont
  {Horton}},\ }\href {\doibase 10.1103/RevModPhys.71.735} {\bibfield  {journal}
  {\bibinfo  {journal} {Rev. Mod. Phys.}\ }\textbf {\bibinfo {volume} {71}},\
  \bibinfo {pages} {735} (\bibinfo {year} {1999})}\BibitemShut {NoStop}%
\bibitem [{\citenamefont {Xiao}\ and\ \citenamefont
  {Lin}(2009)}]{PhysRevLett.103.085004}%
  \BibitemOpen
  \bibfield  {author} {\bibinfo {author} {\bibfnamefont {Y.}~\bibnamefont
  {Xiao}}\ and\ \bibinfo {author} {\bibfnamefont {Z.}~\bibnamefont {Lin}},\
  }\href {\doibase 10.1103/PhysRevLett.103.085004} {\bibfield  {journal}
  {\bibinfo  {journal} {Phys. Rev. Lett.}\ }\textbf {\bibinfo {volume} {103}},\
  \bibinfo {pages} {085004} (\bibinfo {year} {2009})}\BibitemShut {NoStop}%
\bibitem [{\citenamefont {Deng}\ and\ \citenamefont
  {Lin}(2009)}]{deng2009properties}%
  \BibitemOpen
  \bibfield  {author} {\bibinfo {author} {\bibfnamefont {W.}~\bibnamefont
  {Deng}}\ and\ \bibinfo {author} {\bibfnamefont {Z.}~\bibnamefont {Lin}},\
  }\href@noop {} {\bibfield  {journal} {\bibinfo  {journal} {Physics of
  Plasmas}\ }\textbf {\bibinfo {volume} {16}},\ \bibinfo {pages} {102503}
  (\bibinfo {year} {2009})}\BibitemShut {NoStop}%
\bibitem [{\citenamefont {Lin}\ \emph {et~al.}(2009)\citenamefont {Lin},
  \citenamefont {Porkolab}, \citenamefont {Edlund}, \citenamefont {Rost},
  \citenamefont {Fiore}, \citenamefont {Greenwald}, \citenamefont {Lin},
  \citenamefont {Mikkelsen}, \citenamefont {Tsujii},\ and\ \citenamefont
  {Wukitch}}]{lin2009studies}%
  \BibitemOpen
  \bibfield  {author} {\bibinfo {author} {\bibfnamefont {L.}~\bibnamefont
  {Lin}}, \bibinfo {author} {\bibfnamefont {M.}~\bibnamefont {Porkolab}},
  \bibinfo {author} {\bibfnamefont {E.}~\bibnamefont {Edlund}}, \bibinfo
  {author} {\bibfnamefont {J.}~\bibnamefont {Rost}}, \bibinfo {author}
  {\bibfnamefont {C.}~\bibnamefont {Fiore}}, \bibinfo {author} {\bibfnamefont
  {M.}~\bibnamefont {Greenwald}}, \bibinfo {author} {\bibfnamefont
  {Y.}~\bibnamefont {Lin}}, \bibinfo {author} {\bibfnamefont {D.}~\bibnamefont
  {Mikkelsen}}, \bibinfo {author} {\bibfnamefont {N.}~\bibnamefont {Tsujii}}, \
  and\ \bibinfo {author} {\bibfnamefont {S.}~\bibnamefont {Wukitch}},\
  }\href@noop {} {\bibfield  {journal} {\bibinfo  {journal} {Physics of
  plasmas}\ }\textbf {\bibinfo {volume} {16}},\ \bibinfo {pages} {012502}
  (\bibinfo {year} {2009})}\BibitemShut {NoStop}%
\bibitem [{\citenamefont {Diamond}\ \emph {et~al.}(2009)\citenamefont
  {Diamond}, \citenamefont {McDevitt}, \citenamefont {G{\"u}rcan},
  \citenamefont {Hahm}, \citenamefont {Wang}, \citenamefont {Yoon},
  \citenamefont {Holod}, \citenamefont {Lin}, \citenamefont {Naulin},\ and\
  \citenamefont {Singh}}]{diamond2009physics}%
  \BibitemOpen
  \bibfield  {author} {\bibinfo {author} {\bibfnamefont {P.}~\bibnamefont
  {Diamond}}, \bibinfo {author} {\bibfnamefont {C.}~\bibnamefont {McDevitt}},
  \bibinfo {author} {\bibfnamefont {{\"O}.}~\bibnamefont {G{\"u}rcan}},
  \bibinfo {author} {\bibfnamefont {T.}~\bibnamefont {Hahm}}, \bibinfo {author}
  {\bibfnamefont {W.}~\bibnamefont {Wang}}, \bibinfo {author} {\bibfnamefont
  {E.}~\bibnamefont {Yoon}}, \bibinfo {author} {\bibfnamefont {I.}~\bibnamefont
  {Holod}}, \bibinfo {author} {\bibfnamefont {Z.}~\bibnamefont {Lin}}, \bibinfo
  {author} {\bibfnamefont {V.}~\bibnamefont {Naulin}}, \ and\ \bibinfo {author}
  {\bibfnamefont {R.}~\bibnamefont {Singh}},\ }\href@noop {} {\bibfield
  {journal} {\bibinfo  {journal} {Nuclear Fusion}\ }\textbf {\bibinfo {volume}
  {49}},\ \bibinfo {pages} {045002} (\bibinfo {year} {2009})}\BibitemShut
  {NoStop}%
\bibitem [{\citenamefont {Wesson}\ and\ \citenamefont
  {Campbell}(2011)}]{wesson2011tokamaks}%
  \BibitemOpen
  \bibfield  {author} {\bibinfo {author} {\bibfnamefont {J.}~\bibnamefont
  {Wesson}}\ and\ \bibinfo {author} {\bibfnamefont {D.~J.}\ \bibnamefont
  {Campbell}},\ }\href@noop {} {\emph {\bibinfo {title} {Tokamaks}}},\ Vol.\
  \bibinfo {volume} {149}\ (\bibinfo  {publisher} {Oxford university press},\
  \bibinfo {year} {2011})\BibitemShut {NoStop}%
\bibitem [{\citenamefont {Doyle}\ \emph {et~al.}(2007)\citenamefont {Doyle},
  \citenamefont {Houlberg}, \citenamefont {Kamada}, \citenamefont {Mukhovatov},
  \citenamefont {Osborne}, \citenamefont {Polevoi}, \citenamefont {Bateman},
  \citenamefont {Connor}, \citenamefont {Cordey}, \citenamefont {Fujita} \emph
  {et~al.}}]{doyle2007plasma}%
  \BibitemOpen
  \bibfield  {author} {\bibinfo {author} {\bibfnamefont {E.}~\bibnamefont
  {Doyle}}, \bibinfo {author} {\bibfnamefont {W.}~\bibnamefont {Houlberg}},
  \bibinfo {author} {\bibfnamefont {Y.}~\bibnamefont {Kamada}}, \bibinfo
  {author} {\bibfnamefont {V.}~\bibnamefont {Mukhovatov}}, \bibinfo {author}
  {\bibfnamefont {T.}~\bibnamefont {Osborne}}, \bibinfo {author} {\bibfnamefont
  {A.}~\bibnamefont {Polevoi}}, \bibinfo {author} {\bibfnamefont
  {G.}~\bibnamefont {Bateman}}, \bibinfo {author} {\bibfnamefont
  {J.}~\bibnamefont {Connor}}, \bibinfo {author} {\bibfnamefont
  {J.}~\bibnamefont {Cordey}}, \bibinfo {author} {\bibfnamefont
  {T.}~\bibnamefont {Fujita}},  \emph {et~al.},\ }\href@noop {} {\bibfield
  {journal} {\bibinfo  {journal} {Nuclear Fusion}\ }\textbf {\bibinfo {volume}
  {47}},\ \bibinfo {pages} {S18} (\bibinfo {year} {2007})}\BibitemShut
  {NoStop}%
\bibitem [{\citenamefont {Conner}\ and\ \citenamefont
  {Wilson}(1994)}]{conner1994survey}%
  \BibitemOpen
  \bibfield  {author} {\bibinfo {author} {\bibfnamefont {J.}~\bibnamefont
  {Conner}}\ and\ \bibinfo {author} {\bibfnamefont {H.}~\bibnamefont
  {Wilson}},\ }\href@noop {} {\bibfield  {journal} {\bibinfo  {journal} {Plasma
  Physics and Controlled Fusion}\ }\textbf {\bibinfo {volume} {36}},\ \bibinfo
  {pages} {719} (\bibinfo {year} {1994})}\BibitemShut {NoStop}%
\bibitem [{\citenamefont {Dimits}\ \emph {et~al.}(2000)\citenamefont {Dimits},
  \citenamefont {Bateman}, \citenamefont {Beer}, \citenamefont {Cohen},
  \citenamefont {Dorland}, \citenamefont {Hammett}, \citenamefont {Kim},
  \citenamefont {Kinsey}, \citenamefont {Kotschenreuther}, \citenamefont
  {Kritz} \emph {et~al.}}]{dimits2000comparisons}%
  \BibitemOpen
  \bibfield  {author} {\bibinfo {author} {\bibfnamefont {A.~M.}\ \bibnamefont
  {Dimits}}, \bibinfo {author} {\bibfnamefont {G.}~\bibnamefont {Bateman}},
  \bibinfo {author} {\bibfnamefont {M.}~\bibnamefont {Beer}}, \bibinfo {author}
  {\bibfnamefont {B.}~\bibnamefont {Cohen}}, \bibinfo {author} {\bibfnamefont
  {W.}~\bibnamefont {Dorland}}, \bibinfo {author} {\bibfnamefont
  {G.}~\bibnamefont {Hammett}}, \bibinfo {author} {\bibfnamefont
  {C.}~\bibnamefont {Kim}}, \bibinfo {author} {\bibfnamefont {J.}~\bibnamefont
  {Kinsey}}, \bibinfo {author} {\bibfnamefont {M.}~\bibnamefont
  {Kotschenreuther}}, \bibinfo {author} {\bibfnamefont {A.}~\bibnamefont
  {Kritz}},  \emph {et~al.},\ }\href@noop {} {\bibfield  {journal} {\bibinfo
  {journal} {Physics of Plasmas}\ }\textbf {\bibinfo {volume} {7}},\ \bibinfo
  {pages} {969} (\bibinfo {year} {2000})}\BibitemShut {NoStop}%
\bibitem [{\citenamefont {Jenko}\ \emph {et~al.}(2000)\citenamefont {Jenko},
  \citenamefont {Dorland}, \citenamefont {Kotschenreuther},\ and\ \citenamefont
  {Rogers}}]{jenko2000electron}%
  \BibitemOpen
  \bibfield  {author} {\bibinfo {author} {\bibfnamefont {F.}~\bibnamefont
  {Jenko}}, \bibinfo {author} {\bibfnamefont {W.}~\bibnamefont {Dorland}},
  \bibinfo {author} {\bibfnamefont {M.}~\bibnamefont {Kotschenreuther}}, \ and\
  \bibinfo {author} {\bibfnamefont {B.}~\bibnamefont {Rogers}},\ }\href@noop {}
  {\bibfield  {journal} {\bibinfo  {journal} {Physics of Plasmas}\ }\textbf
  {\bibinfo {volume} {7}},\ \bibinfo {pages} {1904} (\bibinfo {year}
  {2000})}\BibitemShut {NoStop}%
\bibitem [{\citenamefont {Lin}\ \emph {et~al.}(1998)\citenamefont {Lin},
  \citenamefont {Hahm}, \citenamefont {Lee}, \citenamefont {Tang},\ and\
  \citenamefont {White}}]{lin1998turbulent}%
  \BibitemOpen
  \bibfield  {author} {\bibinfo {author} {\bibfnamefont {Z.}~\bibnamefont
  {Lin}}, \bibinfo {author} {\bibfnamefont {T.~S.}\ \bibnamefont {Hahm}},
  \bibinfo {author} {\bibfnamefont {W.}~\bibnamefont {Lee}}, \bibinfo {author}
  {\bibfnamefont {W.~M.}\ \bibnamefont {Tang}}, \ and\ \bibinfo {author}
  {\bibfnamefont {R.~B.}\ \bibnamefont {White}},\ }\href@noop {} {\bibfield
  {journal} {\bibinfo  {journal} {Science}\ }\textbf {\bibinfo {volume}
  {281}},\ \bibinfo {pages} {1835} (\bibinfo {year} {1998})}\BibitemShut
  {NoStop}%
\bibitem [{\citenamefont {Chowdhury}\ \emph {et~al.}(2009)\citenamefont
  {Chowdhury}, \citenamefont {Ganesh}, \citenamefont {Brunner}, \citenamefont
  {Vaclavik}, \citenamefont {Villard},\ and\ \citenamefont
  {Angelino}}]{chowdhury2009comprehensive}%
  \BibitemOpen
  \bibfield  {author} {\bibinfo {author} {\bibfnamefont {J.}~\bibnamefont
  {Chowdhury}}, \bibinfo {author} {\bibfnamefont {R.}~\bibnamefont {Ganesh}},
  \bibinfo {author} {\bibfnamefont {S.}~\bibnamefont {Brunner}}, \bibinfo
  {author} {\bibfnamefont {J.}~\bibnamefont {Vaclavik}}, \bibinfo {author}
  {\bibfnamefont {L.}~\bibnamefont {Villard}}, \ and\ \bibinfo {author}
  {\bibfnamefont {P.}~\bibnamefont {Angelino}},\ }\href@noop {} {\bibfield
  {journal} {\bibinfo  {journal} {Physics of Plasmas}\ }\textbf {\bibinfo
  {volume} {16}},\ \bibinfo {pages} {052507} (\bibinfo {year}
  {2009})}\BibitemShut {NoStop}%
\bibitem [{\citenamefont {Gang}\ and\ \citenamefont
  {Diamond}(1990)}]{gang1990nonlinear}%
  \BibitemOpen
  \bibfield  {author} {\bibinfo {author} {\bibfnamefont {F.}~\bibnamefont
  {Gang}}\ and\ \bibinfo {author} {\bibfnamefont {P.}~\bibnamefont {Diamond}},\
  }\href@noop {} {\bibfield  {journal} {\bibinfo  {journal} {Physics of Fluids
  B: Plasma Physics}\ }\textbf {\bibinfo {volume} {2}},\ \bibinfo {pages}
  {2976} (\bibinfo {year} {1990})}\BibitemShut {NoStop}%
\bibitem [{\citenamefont {Gang}, \citenamefont {Diamond},\ and\ \citenamefont
  {Rosenbluth}(1991)}]{gang1991kinetic}%
  \BibitemOpen
  \bibfield  {author} {\bibinfo {author} {\bibfnamefont {F.}~\bibnamefont
  {Gang}}, \bibinfo {author} {\bibfnamefont {P.}~\bibnamefont {Diamond}}, \
  and\ \bibinfo {author} {\bibfnamefont {M.}~\bibnamefont {Rosenbluth}},\
  }\href@noop {} {\bibfield  {journal} {\bibinfo  {journal} {Physics of Fluids
  B: Plasma Physics}\ }\textbf {\bibinfo {volume} {3}},\ \bibinfo {pages} {68}
  (\bibinfo {year} {1991})}\BibitemShut {NoStop}%
\bibitem [{\citenamefont {Hahm}\ and\ \citenamefont
  {Tang}(1991)}]{hahm1991weak}%
  \BibitemOpen
  \bibfield  {author} {\bibinfo {author} {\bibfnamefont {T.}~\bibnamefont
  {Hahm}}\ and\ \bibinfo {author} {\bibfnamefont {W.}~\bibnamefont {Tang}},\
  }\href@noop {} {\bibfield  {journal} {\bibinfo  {journal} {Physics of Fluids
  B: Plasma Physics}\ }\textbf {\bibinfo {volume} {3}},\ \bibinfo {pages} {989}
  (\bibinfo {year} {1991})}\BibitemShut {NoStop}%
\bibitem [{\citenamefont {Hahm}(1991)}]{hahm1991nonlinear}%
  \BibitemOpen
  \bibfield  {author} {\bibinfo {author} {\bibfnamefont {T.~S.}\ \bibnamefont
  {Hahm}},\ }\href@noop {} {\bibfield  {journal} {\bibinfo  {journal} {Physics
  of Fluids B: Plasma Physics}\ }\textbf {\bibinfo {volume} {3}},\ \bibinfo
  {pages} {1445} (\bibinfo {year} {1991})}\BibitemShut {NoStop}%
\bibitem [{\citenamefont {Beer}\ and\ \citenamefont
  {Hammett}(1996)}]{beer1996bounce}%
  \BibitemOpen
  \bibfield  {author} {\bibinfo {author} {\bibfnamefont {M.~A.}\ \bibnamefont
  {Beer}}\ and\ \bibinfo {author} {\bibfnamefont {G.~W.}\ \bibnamefont
  {Hammett}},\ }\href@noop {} {\bibfield  {journal} {\bibinfo  {journal}
  {Physics of Plasmas}\ }\textbf {\bibinfo {volume} {3}},\ \bibinfo {pages}
  {4018} (\bibinfo {year} {1996})}\BibitemShut {NoStop}%
\bibitem [{\citenamefont {Kinsey}, \citenamefont {Waltz},\ and\ \citenamefont
  {Candy}(2006)}]{kinsey2006effect}%
  \BibitemOpen
  \bibfield  {author} {\bibinfo {author} {\bibfnamefont {J.}~\bibnamefont
  {Kinsey}}, \bibinfo {author} {\bibfnamefont {R.}~\bibnamefont {Waltz}}, \
  and\ \bibinfo {author} {\bibfnamefont {J.}~\bibnamefont {Candy}},\
  }\href@noop {} {\bibfield  {journal} {\bibinfo  {journal} {Physics of
  plasmas}\ }\textbf {\bibinfo {volume} {13}},\ \bibinfo {pages} {022305}
  (\bibinfo {year} {2006})}\BibitemShut {NoStop}%
\bibitem [{\citenamefont {Ernst}\ \emph {et~al.}(2004)\citenamefont {Ernst},
  \citenamefont {Bonoli}, \citenamefont {Catto}, \citenamefont {Dorland},
  \citenamefont {Fiore}, \citenamefont {Granetz}, \citenamefont {Greenwald},
  \citenamefont {Hubbard}, \citenamefont {Porkolab}, \citenamefont {Redi} \emph
  {et~al.}}]{ernst2004role}%
  \BibitemOpen
  \bibfield  {author} {\bibinfo {author} {\bibfnamefont {.~D.}\ \bibnamefont
  {Ernst}}, \bibinfo {author} {\bibfnamefont {P.}~\bibnamefont {Bonoli}},
  \bibinfo {author} {\bibfnamefont {P.}~\bibnamefont {Catto}}, \bibinfo
  {author} {\bibfnamefont {W.}~\bibnamefont {Dorland}}, \bibinfo {author}
  {\bibfnamefont {C.}~\bibnamefont {Fiore}}, \bibinfo {author} {\bibfnamefont
  {R.}~\bibnamefont {Granetz}}, \bibinfo {author} {\bibfnamefont
  {M.}~\bibnamefont {Greenwald}}, \bibinfo {author} {\bibfnamefont
  {A.}~\bibnamefont {Hubbard}}, \bibinfo {author} {\bibfnamefont
  {M.}~\bibnamefont {Porkolab}}, \bibinfo {author} {\bibfnamefont
  {M.}~\bibnamefont {Redi}},  \emph {et~al.},\ }\href@noop {} {\bibfield
  {journal} {\bibinfo  {journal} {Physics of Plasmas}\ }\textbf {\bibinfo
  {volume} {11}},\ \bibinfo {pages} {2637} (\bibinfo {year}
  {2004})}\BibitemShut {NoStop}%
\bibitem [{\citenamefont {Ryter}\ \emph {et~al.}(2001)\citenamefont {Ryter},
  \citenamefont {Imbeaux}, \citenamefont {Leuterer}, \citenamefont {Fahrbach},
  \citenamefont {Suttrop},\ and\ \citenamefont {Team}}]{PhysRevLett.86.5498}%
  \BibitemOpen
  \bibfield  {author} {\bibinfo {author} {\bibfnamefont {F.}~\bibnamefont
  {Ryter}}, \bibinfo {author} {\bibfnamefont {F.}~\bibnamefont {Imbeaux}},
  \bibinfo {author} {\bibfnamefont {F.}~\bibnamefont {Leuterer}}, \bibinfo
  {author} {\bibfnamefont {H.-U.}\ \bibnamefont {Fahrbach}}, \bibinfo {author}
  {\bibfnamefont {W.}~\bibnamefont {Suttrop}}, \ and\ \bibinfo {author}
  {\bibfnamefont {A.~U.}\ \bibnamefont {Team}},\ }\href {\doibase
  10.1103/PhysRevLett.86.5498} {\bibfield  {journal} {\bibinfo  {journal}
  {Phys. Rev. Lett.}\ }\textbf {\bibinfo {volume} {86}},\ \bibinfo {pages}
  {5498} (\bibinfo {year} {2001})}\BibitemShut {NoStop}%
\bibitem [{\citenamefont {DeBoo}\ \emph {et~al.}(2005)\citenamefont {DeBoo},
  \citenamefont {Cirant}, \citenamefont {Luce}, \citenamefont {Manini},
  \citenamefont {Petty}, \citenamefont {Ryter}, \citenamefont {Austin},
  \citenamefont {Baker}, \citenamefont {Gentle}, \citenamefont {Greenfield}
  \emph {et~al.}}]{deboo2005search}%
  \BibitemOpen
  \bibfield  {author} {\bibinfo {author} {\bibfnamefont {J.}~\bibnamefont
  {DeBoo}}, \bibinfo {author} {\bibfnamefont {S.}~\bibnamefont {Cirant}},
  \bibinfo {author} {\bibfnamefont {T.}~\bibnamefont {Luce}}, \bibinfo {author}
  {\bibfnamefont {A.}~\bibnamefont {Manini}}, \bibinfo {author} {\bibfnamefont
  {C.}~\bibnamefont {Petty}}, \bibinfo {author} {\bibfnamefont
  {F.}~\bibnamefont {Ryter}}, \bibinfo {author} {\bibfnamefont
  {M.}~\bibnamefont {Austin}}, \bibinfo {author} {\bibfnamefont
  {D.}~\bibnamefont {Baker}}, \bibinfo {author} {\bibfnamefont
  {K.}~\bibnamefont {Gentle}}, \bibinfo {author} {\bibfnamefont
  {C.}~\bibnamefont {Greenfield}},  \emph {et~al.},\ }\href@noop {} {\bibfield
  {journal} {\bibinfo  {journal} {Nuclear fusion}\ }\textbf {\bibinfo {volume}
  {45}},\ \bibinfo {pages} {494} (\bibinfo {year} {2005})}\BibitemShut
  {NoStop}%
\bibitem [{\citenamefont {Ince-Cushman}\ \emph {et~al.}(2009)\citenamefont
  {Ince-Cushman}, \citenamefont {Rice}, \citenamefont {Reinke}, \citenamefont
  {Greenwald}, \citenamefont {Wallace}, \citenamefont {Parker}, \citenamefont
  {Fiore}, \citenamefont {Hughes}, \citenamefont {Bonoli}, \citenamefont
  {Shiraiwa}, \citenamefont {Hubbard}, \citenamefont {Wolfe}, \citenamefont
  {Hutchinson}, \citenamefont {Marmar}, \citenamefont {Bitter}, \citenamefont
  {Wilson},\ and\ \citenamefont {Hill}}]{PhysRevLett.102.035002}%
  \BibitemOpen
  \bibfield  {author} {\bibinfo {author} {\bibfnamefont {A.}~\bibnamefont
  {Ince-Cushman}}, \bibinfo {author} {\bibfnamefont {J.~E.}\ \bibnamefont
  {Rice}}, \bibinfo {author} {\bibfnamefont {M.}~\bibnamefont {Reinke}},
  \bibinfo {author} {\bibfnamefont {M.}~\bibnamefont {Greenwald}}, \bibinfo
  {author} {\bibfnamefont {G.}~\bibnamefont {Wallace}}, \bibinfo {author}
  {\bibfnamefont {R.}~\bibnamefont {Parker}}, \bibinfo {author} {\bibfnamefont
  {C.}~\bibnamefont {Fiore}}, \bibinfo {author} {\bibfnamefont {J.~W.}\
  \bibnamefont {Hughes}}, \bibinfo {author} {\bibfnamefont {P.}~\bibnamefont
  {Bonoli}}, \bibinfo {author} {\bibfnamefont {S.}~\bibnamefont {Shiraiwa}},
  \bibinfo {author} {\bibfnamefont {A.}~\bibnamefont {Hubbard}}, \bibinfo
  {author} {\bibfnamefont {S.}~\bibnamefont {Wolfe}}, \bibinfo {author}
  {\bibfnamefont {I.~H.}\ \bibnamefont {Hutchinson}}, \bibinfo {author}
  {\bibfnamefont {E.}~\bibnamefont {Marmar}}, \bibinfo {author} {\bibfnamefont
  {M.}~\bibnamefont {Bitter}}, \bibinfo {author} {\bibfnamefont
  {J.}~\bibnamefont {Wilson}}, \ and\ \bibinfo {author} {\bibfnamefont
  {K.}~\bibnamefont {Hill}},\ }\href {\doibase 10.1103/PhysRevLett.102.035002}
  {\bibfield  {journal} {\bibinfo  {journal} {Phys. Rev. Lett.}\ }\textbf
  {\bibinfo {volume} {102}},\ \bibinfo {pages} {035002} (\bibinfo {year}
  {2009})}\BibitemShut {NoStop}%
\bibitem [{\citenamefont {Rice}\ \emph {et~al.}(2009)\citenamefont {Rice},
  \citenamefont {Ince-Cushman}, \citenamefont {Bonoli}, \citenamefont
  {Greenwald}, \citenamefont {Hughes}, \citenamefont {Parker}, \citenamefont
  {Reinke}, \citenamefont {Wallace}, \citenamefont {Fiore}, \citenamefont
  {Granetz} \emph {et~al.}}]{rice2009observations}%
  \BibitemOpen
  \bibfield  {author} {\bibinfo {author} {\bibfnamefont {J.}~\bibnamefont
  {Rice}}, \bibinfo {author} {\bibfnamefont {A.}~\bibnamefont {Ince-Cushman}},
  \bibinfo {author} {\bibfnamefont {P.}~\bibnamefont {Bonoli}}, \bibinfo
  {author} {\bibfnamefont {M.}~\bibnamefont {Greenwald}}, \bibinfo {author}
  {\bibfnamefont {J.}~\bibnamefont {Hughes}}, \bibinfo {author} {\bibfnamefont
  {R.}~\bibnamefont {Parker}}, \bibinfo {author} {\bibfnamefont
  {M.}~\bibnamefont {Reinke}}, \bibinfo {author} {\bibfnamefont
  {G.}~\bibnamefont {Wallace}}, \bibinfo {author} {\bibfnamefont
  {C.}~\bibnamefont {Fiore}}, \bibinfo {author} {\bibfnamefont
  {R.}~\bibnamefont {Granetz}},  \emph {et~al.},\ }\href@noop {} {\bibfield
  {journal} {\bibinfo  {journal} {Nuclear Fusion}\ }\textbf {\bibinfo {volume}
  {49}},\ \bibinfo {pages} {025004} (\bibinfo {year} {2009})}\BibitemShut
  {NoStop}%
\bibitem [{\citenamefont {Liu}, \citenamefont {Baxter},\ and\ \citenamefont
  {Thompson}(1971)}]{PhysRevLett.26.621}%
  \BibitemOpen
  \bibfield  {author} {\bibinfo {author} {\bibfnamefont {C.~S.}\ \bibnamefont
  {Liu}}, \bibinfo {author} {\bibfnamefont {D.~C.}\ \bibnamefont {Baxter}}, \
  and\ \bibinfo {author} {\bibfnamefont {W.~B.}\ \bibnamefont {Thompson}},\
  }\href {\doibase 10.1103/PhysRevLett.26.621} {\bibfield  {journal} {\bibinfo
  {journal} {Phys. Rev. Lett.}\ }\textbf {\bibinfo {volume} {26}},\ \bibinfo
  {pages} {621} (\bibinfo {year} {1971})}\BibitemShut {NoStop}%
\bibitem [{\citenamefont {Liu}\ and\ \citenamefont
  {Tripathi}(1980)}]{liu1980stabilization}%
  \BibitemOpen
  \bibfield  {author} {\bibinfo {author} {\bibfnamefont {C.}~\bibnamefont
  {Liu}}\ and\ \bibinfo {author} {\bibfnamefont {V.}~\bibnamefont {Tripathi}},\
  }\href@noop {} {\bibfield  {journal} {\bibinfo  {journal} {The Physics of
  Fluids}\ }\textbf {\bibinfo {volume} {23}},\ \bibinfo {pages} {345} (\bibinfo
  {year} {1980})}\BibitemShut {NoStop}%
\bibitem [{\citenamefont {Praburam}, \citenamefont {Tripathi},\ and\
  \citenamefont {Jain}(1988)}]{praburam1988lower}%
  \BibitemOpen
  \bibfield  {author} {\bibinfo {author} {\bibfnamefont {G.}~\bibnamefont
  {Praburam}}, \bibinfo {author} {\bibfnamefont {V.}~\bibnamefont {Tripathi}},
  \ and\ \bibinfo {author} {\bibfnamefont {V.}~\bibnamefont {Jain}},\
  }\href@noop {} {\bibfield  {journal} {\bibinfo  {journal} {The Physics of
  fluids}\ }\textbf {\bibinfo {volume} {31}},\ \bibinfo {pages} {3145}
  (\bibinfo {year} {1988})}\BibitemShut {NoStop}%
\bibitem [{\citenamefont {Wong}\ and\ \citenamefont
  {Bellan}(1978)}]{wong1978enhancement}%
  \BibitemOpen
  \bibfield  {author} {\bibinfo {author} {\bibfnamefont {K.}~\bibnamefont
  {Wong}}\ and\ \bibinfo {author} {\bibfnamefont {P.}~\bibnamefont {Bellan}},\
  }\href@noop {} {\bibfield  {journal} {\bibinfo  {journal} {The Physics of
  Fluids}\ }\textbf {\bibinfo {volume} {21}},\ \bibinfo {pages} {841} (\bibinfo
  {year} {1978})}\BibitemShut {NoStop}%
\bibitem [{\citenamefont {Redi}\ \emph {et~al.}(2005)\citenamefont {Redi},
  \citenamefont {Dorland}, \citenamefont {Fiore}, \citenamefont {Baumgaertel},
  \citenamefont {Belli}, \citenamefont {Hahm}, \citenamefont {Hammett},\ and\
  \citenamefont {Rewoldt}}]{redi2005microturbulent}%
  \BibitemOpen
  \bibfield  {author} {\bibinfo {author} {\bibfnamefont {M.}~\bibnamefont
  {Redi}}, \bibinfo {author} {\bibfnamefont {W.}~\bibnamefont {Dorland}},
  \bibinfo {author} {\bibfnamefont {C.}~\bibnamefont {Fiore}}, \bibinfo
  {author} {\bibfnamefont {J.}~\bibnamefont {Baumgaertel}}, \bibinfo {author}
  {\bibfnamefont {E.}~\bibnamefont {Belli}}, \bibinfo {author} {\bibfnamefont
  {T.}~\bibnamefont {Hahm}}, \bibinfo {author} {\bibfnamefont {G.}~\bibnamefont
  {Hammett}}, \ and\ \bibinfo {author} {\bibfnamefont {G.}~\bibnamefont
  {Rewoldt}},\ }\href@noop {} {\bibfield  {journal} {\bibinfo  {journal}
  {Physics of plasmas}\ }\textbf {\bibinfo {volume} {12}},\ \bibinfo {pages}
  {072519} (\bibinfo {year} {2005})}\BibitemShut {NoStop}%
\bibitem [{\citenamefont {Kuley}\ and\ \citenamefont
  {Tripathi}(2009)}]{kuley2009stabilization}%
  \BibitemOpen
  \bibfield  {author} {\bibinfo {author} {\bibfnamefont {A.}~\bibnamefont
  {Kuley}}\ and\ \bibinfo {author} {\bibfnamefont {V.}~\bibnamefont
  {Tripathi}},\ }\href@noop {} {\bibfield  {journal} {\bibinfo  {journal}
  {Physics of Plasmas}\ }\textbf {\bibinfo {volume} {16}},\ \bibinfo {pages}
  {032504} (\bibinfo {year} {2009})}\BibitemShut {NoStop}%
\bibitem [{\citenamefont {Connor}, \citenamefont {Hastie},\ and\ \citenamefont
  {Helander}(2006)}]{connor2006stability}%
  \BibitemOpen
  \bibfield  {author} {\bibinfo {author} {\bibfnamefont {J.}~\bibnamefont
  {Connor}}, \bibinfo {author} {\bibfnamefont {R.}~\bibnamefont {Hastie}}, \
  and\ \bibinfo {author} {\bibfnamefont {P.}~\bibnamefont {Helander}},\
  }\href@noop {} {\bibfield  {journal} {\bibinfo  {journal} {Plasma physics and
  controlled fusion}\ }\textbf {\bibinfo {volume} {48}},\ \bibinfo {pages}
  {885} (\bibinfo {year} {2006})}\BibitemShut {NoStop}%
\bibitem [{\citenamefont {Liu}\ \emph {et~al.}(1984)\citenamefont {Liu},
  \citenamefont {Tripathi}, \citenamefont {Chan},\ and\ \citenamefont
  {Stefan}}]{liu1984density}%
  \BibitemOpen
  \bibfield  {author} {\bibinfo {author} {\bibfnamefont {C.}~\bibnamefont
  {Liu}}, \bibinfo {author} {\bibfnamefont {V.}~\bibnamefont {Tripathi}},
  \bibinfo {author} {\bibfnamefont {V.}~\bibnamefont {Chan}}, \ and\ \bibinfo
  {author} {\bibfnamefont {V.}~\bibnamefont {Stefan}},\ }\href@noop {}
  {\bibfield  {journal} {\bibinfo  {journal} {The Physics of fluids}\ }\textbf
  {\bibinfo {volume} {27}},\ \bibinfo {pages} {1709} (\bibinfo {year}
  {1984})}\BibitemShut {NoStop}%
\bibitem [{\citenamefont {Ware}(1970)}]{PhysRevLett.25.15}%
  \BibitemOpen
  \bibfield  {author} {\bibinfo {author} {\bibfnamefont {A.~A.}\ \bibnamefont
  {Ware}},\ }\href {\doibase 10.1103/PhysRevLett.25.15} {\bibfield  {journal}
  {\bibinfo  {journal} {Phys. Rev. Lett.}\ }\textbf {\bibinfo {volume} {25}},\
  \bibinfo {pages} {15} (\bibinfo {year} {1970})}\BibitemShut {NoStop}%
\end{thebibliography}%

\end{document}